# Simultaneous Orientational and Conformational Molecular Dynamics in Solid (1,1,2)-Trichloroethane

*Michela Romanini,[1,] Efstratia Mitsari,[1] Pragya Tripathi,[1,†] Pablo Serra,[2] Mariano Zuriaga,[2] Josep Lluis Tamarit,[1] and Roberto Macovez [1,\*]*

[1] Grup de Caracterització de Materials, Departament de Física and Barcelona Research Center in Multiscale Science and Engineering, Universitat Politècnica de Catalunya, EEBE, Campus Diagonal-Besòs, Av. Eduard Maristany 10-14, 08019 Barcelona, Catalonia, Spain.

[2] Facultad de Matemática, Astronomía, Física y Computación, Universidad Nacional de Córdoba and IFEG-CONICET, Ciudad Universitaria, X5016LAE Córdoba, Argentina

\* Author to whom correspondence should be addressed. Electronic mail: roberto.macovez@upc.edu

ABSTRACT. The molecular dynamics in the ambient-pressure solid phase of (1,1,2)-trichloroethane is studied by means of broadband dielectric spectroscopy and molecular dynamics simulations. The dielectric spectra of polycrystalline samples obtained by crystallization from the liquid phase exhibit, besides a space-charge relaxation associated with accumulation of charges at crystalline domain boundaries, two loss features arising from dipolar molecular relaxations. The most prominent and slower of the two loss features is identified as a configurational leap of the molecules which involves a simultaneous change in spatial orientation and structural conformation, namely between two isomeric forms (gauche$^+$ and gauche$^-$) of opposite chirality. In this peculiar dynamic process, the positions of the three chlorine atoms in the crystal lattice remain unchanged, while those of the carbon and hydrogen atoms are modified. This dynamic process is responsible for the disorder observed in an earlier x-ray diffraction study and confirmed by our simulation, which is present only at temperatures relatively close to the melting point, starting 40 K below. The onset of the disorder is visible as an anomaly in the temperature dependence of the dc conductivity of the sample at exactly the same temperature. While the slower relaxation dynamics (combined isomerization/reorientation) becomes increasingly more intense on approaching the melting point, the faster dynamics exhibits significantly lower but constant dielectric strength. Based on our molecular dynamics simulations, we assign the faster relaxation to large fluctuations of the molecular dipole moments, partly due to large-angle librations of the chloroethane species.

Introduction

Ethane derivatives are the smallest organic molecules undergoing conformational conversions which modify the molecular chirality. Modified ethanes are therefore the simplest systems where properties such as dynamic chiral isomerism and torsional flexibility can be investigated. In this respect, halogenated ethanes are a particularly interesting model system. Compared to other ethane derivatives such as ethanol, ethylamine or ethanolamine, for example, the intermolecular coupling and molecular dynamics of halogenated ethanes are expected to be simpler due to the lack of directional hydrogen-bond interactions, so that self-aggregation and networking are not expected to occur. Nonetheless, due to their delicate conformational equilibrium, halogenated ethanes such as chloroethanes display very fascinating and rich conformational phase diagrams,[1-9] which include in some cases dynamic orientational or conformational disorder.[10-15] While some chloroethanes exhibit, as pristine ethane, only one possible conformation (*e.g.* monochloroethane), others, such as (1,2)-dichloroethane, (1,1,2)-trichloroethane, and (1,1,2,2)-tetrachloroethane, exhibit three different conformations, namely *gauche⁺*, *gauche⁻* and *transoid*, depending on the relative positions of the chlorines linked to either carbon atom (in the *transoid* isomer of (1,1,2,2)-tetrachloroethane, it is actually the hydrogen on carbon 1 and the chlorine on carbon 2 that are in a *trans* configuration with respect to each another). The occurrence and relative concentration of conformers varies according to the nature of the system, and depends on the subtle balance of intramolecular strains and the effects of the molecular environment.[14,16-18]

We focus here on (1,1,2)-trichloroethane (hereafter TCE), of molecular formula $C_2H_3Cl_3$. At ambient pressure, TCE is liquid at room temperature and below 237 K it displays a monoclinic phase, referred to as phase α, characterized by the simultaneous presence of distinct molecular conformers and orientations, accompanied by site disorder close to the melting point.[14] At high pressure, a different monoclinic solid phase is observed (referred to as phase *β*), which at room temperature is stable above 0.82 GPa. Different conformers are present in the different phases of TCE. The *gauche* conformer is energetically more stable in isolation and by far the most abundant in the gas phase,[19,20] where intermolecular interactions are negligible. Vibrational and NMR studies on the pure solvent have shown instead that in the liquid phase the $C_2H_3Cl_3$ molecules exist in both *gauche* and *transoid* conformations, with comparable concentrations.[21-24] In the *β* crystalline phase (at high pressure) only *transoid* conformers are observed. The α solid phase is arguably the most interesting one: only *gauche* conformers are present, but of both chirality, namely *gauche*$^+$ and *gauche*$^-$. The simultaneous presence of two conformers has been reported also in 1:1 stoichiometric cocrystals of (1,1,2)-trichloroethane with Buckminsterfullerene.[18]

The disorder in phase α is particularly intriguing. While at low temperature the *gauche*$^+$ and *gauche*$^-$ conformers each occupy distinct crystallographic sites related by an inversion center, close to the melting point of this phase (237 K) all sites exhibit orientational and conformational disorder, with the molecules in one of two possible configurations. At 220 K, the observed occupancies of the majority and minority conformer at each site are 0.85 and 0.15, respectively.[14] An interesting aspect of this disorder is that the bulky halogen atoms occupy always the same crystallographic positions, while the carbon positions, C–C orientation, and molecular chirality (conformation), change between the two orientations. A similar type of disorder, with the halogens occupying the same lattice position and disorder in the backbone orientation, has been observed also in other halogenated ethanes.[6,7] The authors of Ref. 14 observe a gradual reduction of this disorder as the temperature is lowered, and suggest that this may indicate a dynamic nature of such disorder, that is, that the ethane derivatives may undergo dynamic conformational changes. For this to happen, however, the TCE molecules should undergo a simultaneous orientational and configurational jump between the two different chiral states of the *gauche* conformation.

In this contribution, we employ broadband dielectric spectroscopy and molecular dynamic simulations to investigate the reorientational motions in the solid α phase of TCE. Our dielectric study shows that the peculiar disorder observed in Ref. 14 has indeed a dynamic character. We moreover detect a less intense relaxation process at lower temperature in the same phase. Based on our molecular dynamics simulations, we show that the main dynamic process is a simultaneous orientational-conformational jump of the TCE molecules involving *gauche*$^+$ – *gauche*$^-$ isomerization, and assign the other relaxation to short-lived fluctuations into non-equilibrium configurations, partially ascribable to large-angle librations of the molecules. The presence of two relaxations was reported also in the disordered solid phase of the related (1,1,2,2)-tetrachloroethane derivative,[25] but neither involved a simultaneous change in conformational and orientational degrees of freedom. Conformational molecular dynamics are not uncommon in gaseous or liquid molecular phases, and they have also been also reported in some plastic crystals.[26-28] However, the α phase of TCE is, to the best of our knowledge, the only ordered small-molecule crystal phase that exhibits such kind of dynamic isomerism, which moreover involves a change in chirality.

Materials and Methods

(1,1,2)-Trichloroethane (TCE) was purchased from a commercial supplier (Aldrich, 98%) and distilled twice at 385 K. For the dielectric measurements, the distilled TCE was inserted in its liquid phase inside a home-made stainless steel parallel-plate capacitor especially designed for liquid samples, with the two plates separated by needle-like cylindrical silica spacers of 50 μm diameter. The capacitor was then loaded within a nitrogen-gas flow Quatro cryostat for temperature control. The α crystal phase was obtained by slow cooling of the sample directly in the capacitor cell, and isothermal measurements were taken both upon decreasing and increasing the temperature. Isothermal dielectric spectra were acquired using a Novocontrol Alpha analyzer in the frequency ($f$) range between $10^{-2}$ and $5 \cdot 10^6$ Hz. Dielectric measurements yield the complex impedance of the sample, from which the complex relative permittivity $\varepsilon^*(\omega)$, complex dielectric modulus $M^*(\omega) = \varepsilon^*(\omega)^{-1}$, and complex conductivity $\sigma^*(\omega) = i\omega\varepsilon_0(1 - \varepsilon^*)$ of the material can be retrieved. These frequency-dependent quantities carry information on the dipolar molecular dynamics processes as well as about the dc conductivity and space-charge relaxations of the sample. In particular, dipolar and conductivity-related features can be observed either in the imaginary part of the permittivity $\varepsilon''(f)$, called loss spectrum, or in the imaginary part of the modulus $M''(f)$, called modulus spectrum ($f = \omega/2\pi$).

The value of dc conductivity, $\sigma_{dc}$, was obtained as the low-frequency plateau value of the ac conductivity spectrum, given by $\sigma'(f) = 2\pi f\, \varepsilon_0\, \varepsilon''(f)$. We found that the loss spectra are dominated at low frequency by a conductivity-related space-charge loss, which made impossible a direct fit of the loss spectra. Instead, in the modulus spectra three clearly separated contributions are observed. We therefore determined the frequency maxima of each component from a fit of the modulus spectrum. For this purpose, each component was modeled as a separate peak described with a Havriliak-Negami (HN) function, whose analytical expression in the modulus representation is:[29,30]

(Eq. 1) $M_{HN}(f) = \dfrac{A}{(1+(i2\pi f \tau_{HN})^\beta)^\gamma}$ .

Here, *A* represents the intensity of the process, and the exponents *β* and *γ* are shape parameters related to the low- and high-frequency tails of the imaginary spectrum $M''(f)$, and $\tau_{HN}$ is a fitting parameter from which the characteristic time $\tau_{max}$ at which the imaginary part of the modulus is maximum is obtained as:

(Eq. 2) $\tau_{max} = \tau_{HN} \left(\sin\frac{\beta\pi}{2+2\gamma}\right)^{-1/\beta} \left(\sin\frac{\beta\gamma\pi}{2+2\gamma}\right)^{1/\beta}$.

Two of the components turned out to be characterized by a symmetric peak and well described by either a Cole-Cole and a Cole-Davidson function, which are special cases of HN functions with *γ* = 1 (Cole-Cole) and *β* = 1 (Cole-Davidson), respectively. The obtained values of $\tau_{max,M}$ for all three components can then be analyzed and their temperature dependence studied.

Starting from this knowledge of the relaxation times of all three relaxations, we were able to reproduce the loss spectra $\varepsilon''(f)$ as the imaginary part of the following Eq. (3), which takes into account the presence of three relaxation processes as well as of a conductivity background:

(Eq. 3) $\varepsilon^*(f) = -i\left(\frac{\sigma_0}{\varepsilon_0 2\pi f}\right)^n + \left[\frac{\Delta\varepsilon_{HN}}{[1+(i2\pi f\tau_{HN})^\beta]^\gamma} + \varepsilon_{\infty,HN}\right] + \left[\frac{\Delta\varepsilon_{CC}}{1+(i2\pi f\tau_{CC})^\kappa} + \varepsilon_{\infty,CC}\right] + \left[\frac{\Delta\varepsilon_{CD}}{(1+i2\pi f\tau_{CD})^\delta} + \varepsilon_{\infty,CD}\right]$

The parameter Δε is the dielectric strength (intensity) of each process.

Molecular dynamics simulations of (1,1,2)-trichloroethane were performed using the Gromacs v5.0.2 package[31] in the NPT ensemble, using a system of 960 molecules. Total run times were between 100 to 1000 nanoseconds. Flexible molecules were considered, including harmonic atom-atom forces, harmonic angle potentials and dihedral potential,[32,33] together with a time with time step of 0.5 fs, to allow for conformational changes. The intermolecular interactions were described by Lennard-Jones (L-J) and Coulombic potentials,[32-36] with parameters deduced from

liquid- and gas-phase properties obtained from Ref. 32. As initial configuration, we used the experimental volume and the crystalline structure of the perfectly ordered α phase at 100 K, as determined by X-ray diffraction.[14] In order to allow a comparison with dielectric experiments, we have determined the time self-correlation function of the molecular dipole moment *p* of single molecule, defined as:

(Eq. 4) $C(t) = \langle \vec{p}(t) \cdot \vec{p}(0) \rangle = \frac{1}{N} \sum_i^N \langle \vec{p}_i(\zeta) \cdot \vec{p}_i(t+\zeta) \rangle_\zeta$.

Here N is the number of molecules considered in the calculation, *i* is the molecule number, and the average is carried out over times $\zeta$. The dipole moment correlation function was found to vary slowly, so that extremely long simulation times would be needed to obtain a reliable fit of the correlation decay. For this reason, we also calculated the bond-orientation self-correlation functions given by:

(Eq. 5) $C_j(t) = \langle \vec{b}_j(t) \cdot \vec{b}_j(0) \rangle$.

Here the index *j* indicates a particular C–Cl bond of the TCE molecule. Obviously, any molecular dynamics such as a reorientation motion or a conformational change involves the simultaneous change of the C–C and C–Cl bond directions, so that the time dependence of the latter contains all the information about molecular dynamic processes. The so-obtained self-correlation functions were then fitted as the sum of two exponential decays to mimic the effect of the two dipolar relaxations observed in the modulus spectra. The comparison is meaningful because the dipole time correlation function and the imaginary part of the permittivity are related by the fluctuation-dissipation theorem of linear response theory.

Results and Discussion

In the molecular dynamic simulation, the solid phase α is observed to be stable up to approximately 240 K, *i.e.*, only few degrees higher than the experimental melting point $T_m$ (237 K). Our simulations show that the populations of *gauche*$^+$ and *gauche*$^-$ conformers in phase α are both of 50%, independent of temperature. At low enough temperature, namely, for temperatures lower than $T_m$ – 30 K, each conformer occupies two distinct crystallographic sites in the unit cell (which contains four molecules). The two sites for the *gauche*$^+$ conformer and those for the *gauche*$^-$ conformer are related by an inversion operation.[14] As the temperature is increased from $T_m$ – 30 K, the sites that are initially occupied only by *gauche*$^+$ conformers become partially occupied by *gauche*$^-$ ones and viceversa, in agreement with an earlier x-ray diffraction study.[14]

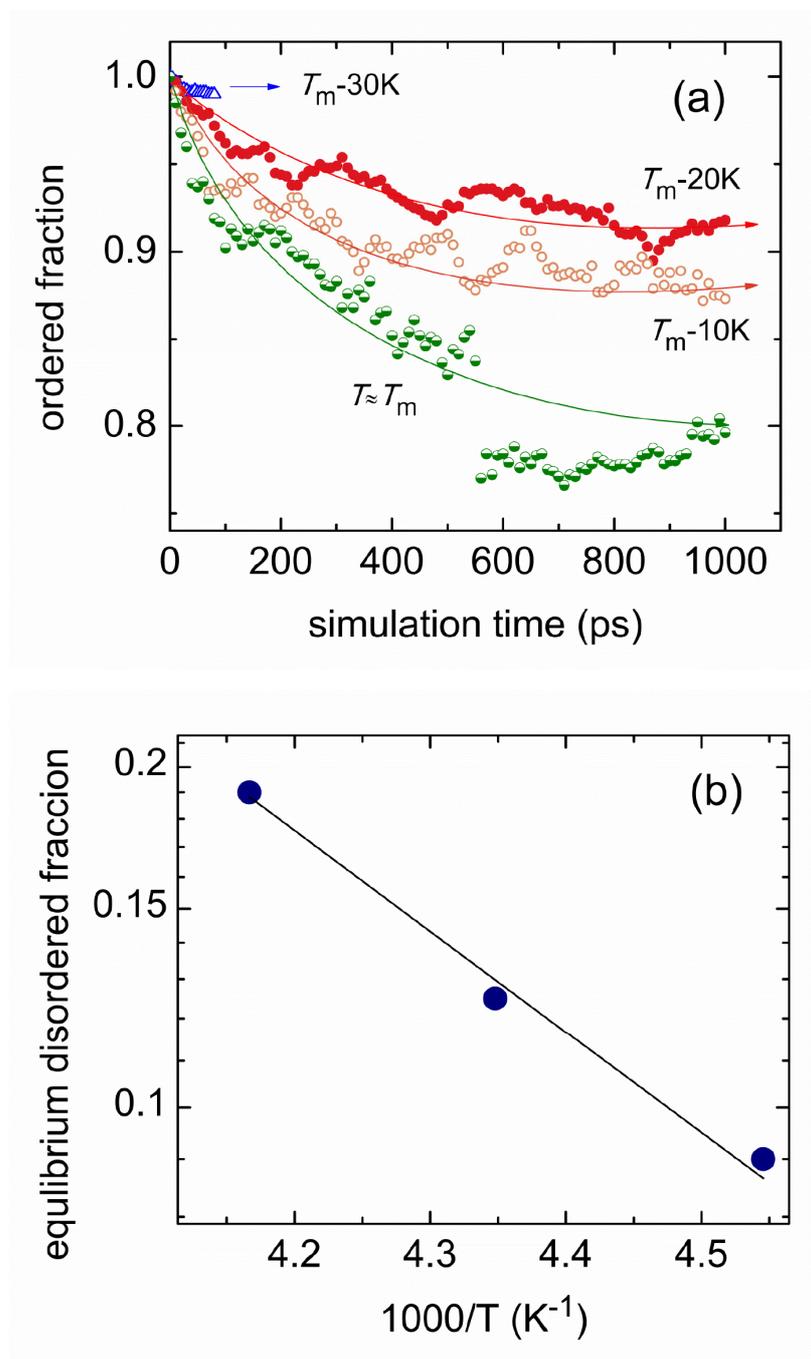

**Figure 1.** Molecular dynamics results imposing as initial condition the equilibrium orientation at low temperature, as obtained from Ref. 14. (a) Evolution with simulation time of the fraction of molecules that maintain the initial orientation (corresponding to the low-temperature equilibrium orientation) at a given site, for different temperatures (referenced to $T_m$). The initial condition is that the fraction is 1 at the beginning of the simulation. (b) Steady-state value, as a function of 1000/T, of the fraction of "disordered" molecules, *i.e.*, molecules that have a different final orientation in the simulation at long simulation times (1000 ns), where a quasi-steady state is reached. The continuous line is a linear fit.

In these simulation runs the initial configuration reproduced the equilibrium situation at low temperature, where all molecules are in the majority orientation at each site and conformational disorder is absent. The system is then allowed to evolve freely from this starting condition. Figure 1(a) shows, as a function of simulation time and for four different temperatures, the fraction of sites that remain occupied by the same conformer that occupies them in the starting condition (i.e., the fraction of sites that are still occupied by the low-T equilibrium conformer). It is seen that for simulation temperatures of $T_\mathrm{m}$ – 30 K, the fraction of sites occupied by the majority conformer is virtually unchanged from its initial value of 1. For higher simulation temperatures (approaching the melting point), it is observed that after an initial transient the same fraction is reduced from 1 to a lower value, which remains more or less stationary at long simulation times. Figure 1(b) shows, for the highest simulation temperatures, the final fraction of "disordered" molecules in the new equilibrium configuration, *i.e.*, the steady-state fraction of molecules that have changed their initial orientation. The number of steady-state disordered molecules increases exponentially with temperature, indicating an activated population of non-equilibrium conformers at a given site. Close to the melting temperature ($T_\mathrm{m}$), this fraction is as high as 20%. At $T_\mathrm{m}$ – 10 K it is close to 15%, in agreement with Ref. 14. It is interesting to note that, although the interconversion rate and the population of a given site depend on temperature, the *total* percentages of each conformer (*gauche*$^+$ and *gauche*$^-$) in the simulation always remain equal to 50%. Assuming that the population of the minority conformers follows an activated temperature dependence, as suggested by the linear fit displayed in Figure 1(b), an activation energy of ~ 4.1 kcal/mol is obtained. This value is close to the calculated potential barrier between *gauche*$^+$ and *gauche*$^-$ conformers in the gas phase (between 4 and 5 kcal/mol depending on the calculation level[19]) and of the same order of magnitude as the potential barrier in solution (between 3 and 7 kcal/mol depending on the solvent and the method used to estimate the energy barrier[19,37]). This comparison suggests that the rotational barrier between the energetically equivalent *gauche*$^+$ and *gauche*$^-$ conformations is mainly determined by intramolecular effects, in contrast with the energy and thus the relative stability of the *gauche* and *transoid* conformers, which are instead strongly dependent on the environment, as mentioned in the introduction.

We next turn to our dielectric spectroscopy data acquired on the α phase of TCE upon heating. Panel (a) of Figure 2 displays the isothermal dielectric loss spectra (ε"(*f*)) at selected temperatures between 115 and 235 K (just below the melting point). The corresponding modulus and ac conductivity spectra are shown in panels (b) and (c), respectively. The σ' spectra display at low frequency a dc conductivity plateau, which corresponds to the linear decrease towards low frequency in the ε" spectra of panel (a). Two separate spectral features can be clearly discerned in the loss spectra, namely, a low-intensity peak visible at low-temperature, labeled as process II, and a more intense loss feature that becomes clearly visible only at high temperature, close to the melting point, and labeled as process I. Besides these two processes, the loss spectra actually comprise also a third loss, labeled as "pre-peak". To highlight the presence of such pre-peak, in the inset to Figure 2(a) we show the so-called derivative loss spectrum, defined as – (π/2) dε'/d(Log*f*),[38] for the permittivity data at 224 K. The derivation procedure allows enhancing the visibility of the dipolar loss features. The maximum of the modulus (panel (b)) corresponds to the so-called "conductivity loss" peak,[39] and it can be seen in the inset to panel (a) that its frequency position matches roughly that of the pre-peak in the permittivity representation. Besides this contribution, two more features are observed in the modulus, at frequencies that match those of processes I and II in the ε" spectra.

In panels (a), (b) and (c) of Figure 2, markers represent experimental spectra and solid lines are fits. In (b), solid lines are fits of the modulus spectra as the sum of three contributions, each described as the imaginary part of a HN function (Eq. (1)). While process II exhibited both shape parameters (HN exponents) different from 1, the conductivity loss could be modeled as a Cole-Cole function and process I with a Cole-Davidson function (see Methods Section). In panels (a) and (c), continuous lines are fits of the loss and ac spectra using Eq. (3) and maintaining the same relaxation times found in the modulus fits.

The plateau value of the σ', which is the dc limit of the conductivity, $\sigma_{dc}$, is displayed as Arrhenius plot in Figure 2(d). The temperature dependence of $\sigma_{dc}$ is simply-activated (Arrhenius) and monotonic up to a temperature of 200 K, or about 40 degrees below the melting point. At this temperature, an anomalous behavior is observed, with virtually no variation of the conductivity over a *T* span of ten degrees, followed by a recovery of the expected monotonic increase all the way up to the melting point. The temperature range of the anomalous behavior of the dc conductivity coincides perfectly with the onset of the molecular disorder reported in Figure 1. It is natural to expect that the onset of disorder hinders the diffusion or hopping of charge carriers,[40] thus counter-acting the beneficial effect of the increased temperature on the activated charge transport process. Hence the $\sigma_{dc}$ anomaly is an indirect confirmation of the results of Figure 1 and of Ref. 14. For comparison purposes, the Arrhenius plot of the relaxation time of the modulus maximum ($\tau_{max,M}$) is also shown in Figure 2(d) (right axis), where it is seen that it exhibits the same temperature dependence as $\sigma_{dc}$ (or more accurately, as the resistivity $1/\sigma_{dc}$). This confirms the identification of the M"($f$) maximum as the conductivity loss, and thus of the pre-peak in ε"($f$), which is visible at almost the same frequency, as a conductivity-related space-charge relaxation. While the peak in the modulus representation arises mostly from the $\sigma_{dc}$ contribution, that is, from the drift of charge carriers under an applied quasi-dc field, in the permittivity representation the $\sigma_{dc}$ contribution is the linear background, while the weak loss feature (pre-peak) is actually a space-charge relaxation due to the accumulation of charge carriers at sample's heterogeneities such as grain boundaries.[41]

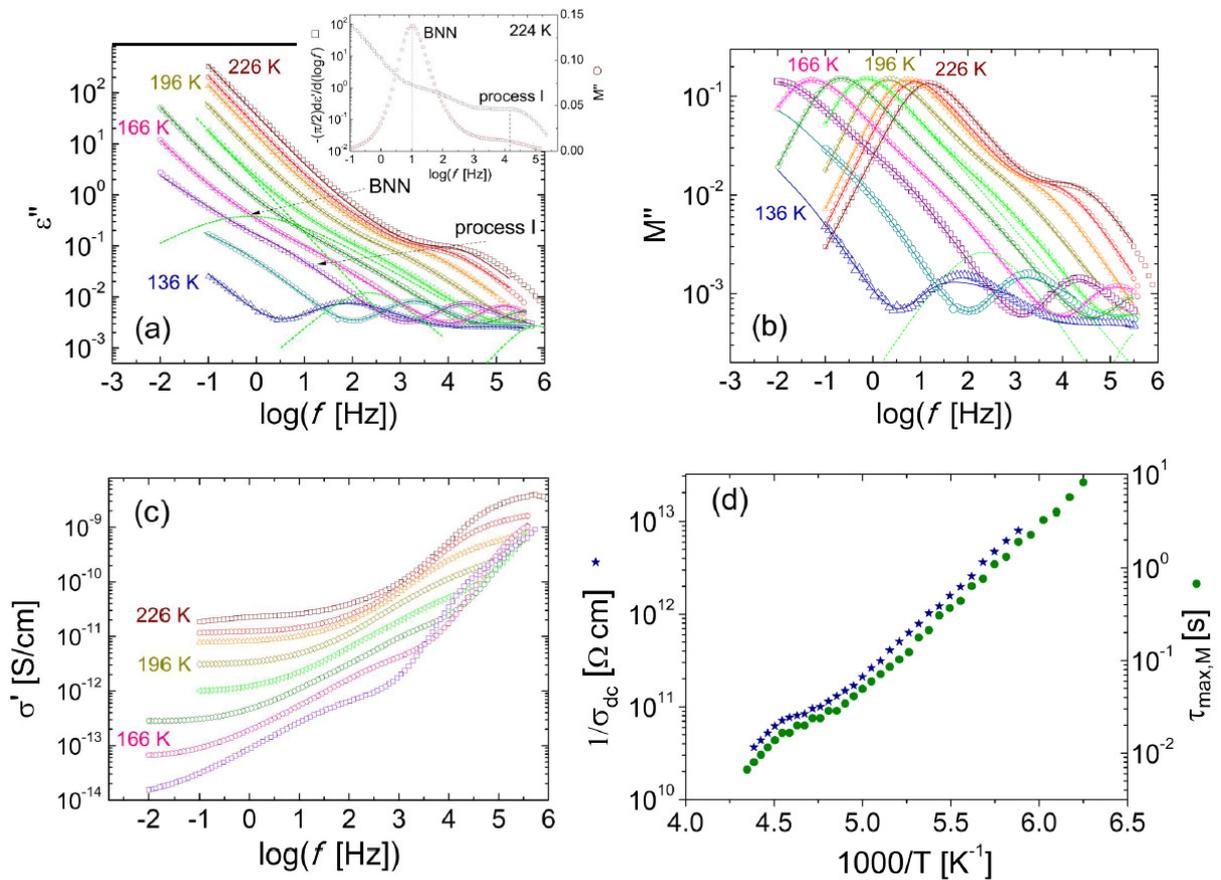

**Figure 2.** Loss ε"($f$) (a), modulus M"($f$) (b), and ac conductivity σ'($f$) (c) isothermal spectra of the solid α phase of TCE, at selected temperatures. Markers are experimental spectra and solid lines are models or fits, as explained in the text. The dotted lines represent the various spectral

components for the spectra acquired at 186 K. Inset to (a): derivative loss spectrum at 224 K obtained from the real permittivity (see the text), and corresponding modulus spectrum. (d) Arrhenius plot of the dc resistivity of TCE, compared with the relaxation time $\tau_{max,M}$ of the most prominent feature in the modulus spectra.

Having discussed the conductivity-related features, we now focus on relaxations I and II, which stem from molecular dynamic processes. Figure 3(a) shows the Arrhenius plot of the relaxation times $\tau_{max,\varepsilon}$ of both relaxations, obtained from the fit of the loss spectra by means of Eq. (3). For comparison, the value of $\tau_{max,M}$ of process II is also shown, as obtained from the fit of the corresponding portion of the modulus spectra with Eq. (1). It may be observed that for process II the values of $\tau_{max,M}$ and $\tau_{max,\varepsilon}$ are virtually identical, as expected because the two quantities are related[42] as $\tau_{max,\varepsilon}/\tau_{max,M} = \varepsilon_s/\varepsilon_\infty$ (where $\varepsilon_s$ and $\varepsilon_\infty$ are respectively the low- and high-frequency limits of the real part of the permittivity above and below the characteristic frequency of relaxation II), and such ratio is close to one because the dielectric strength $\Delta\varepsilon$ of relaxation II, which is equal to the difference $\varepsilon_s - \varepsilon_\infty$, is relatively low. For the same reason, the ratio $\tau_{max,\varepsilon}/\tau_{max,M}$ for relaxation I may be expected to deviate from 1 because of the larger dielectric strength of relaxation I; nonetheless, for simplicity in our model of the loss spectra with Eq. (3) we took $\tau_{max,\varepsilon} = \tau_{max,M}$ at all temperatures for relaxation I.

The inset to Figure 3(a) shows the dielectric strength $\Delta\varepsilon_I$ of process I close to the melting point. The strength of relaxation I exhibits an increase up to the melting temperature, where the relaxation ceases to exist. Such increase is consistent with the observed behavior of the raw loss spectra (Figure 2(a)), where process I is observed to be smeared out below 180 K while it becomes increasingly visible with increasing temperature. In contrast, the dielectric strength of process II decreases slowly with temperature, as it is normally expected for a dipolar relaxation (not shown). The increase of relaxation I in the modulus representation is reminiscent of the increase of disordered sites discussed in relation with Figure 1(b), and matches the observations of an earlier x-ray diffraction study of solid TCE,[14] which reported dynamic orientational disorder near the melting point. The substantial increase of Δε for process I near the melting temperature is a strong indication that it stems from a molecular dynamic process associated with the conformational site-occupancy disorder. Since such disorder involves conformational changes between *gauche$^+$* and *gauche$^-$* isomers with distinct spatial orientation of the C–C bond, we are lead to assign process I to a simultaneous orientational and conformational motion of the TCE molecules. This assignment is corroborated by our molecular dynamics study, as detailed in the following.

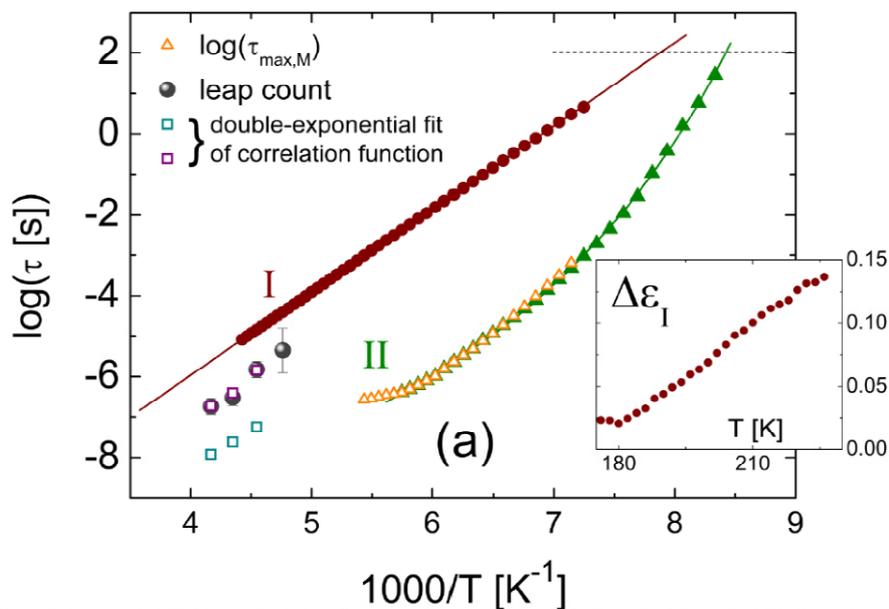

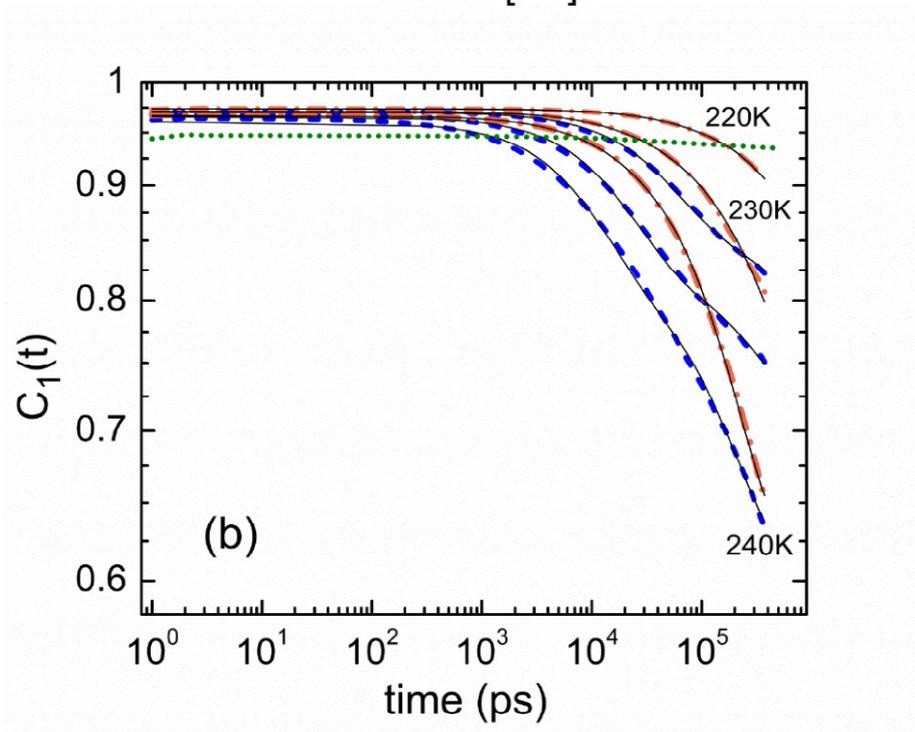

**Figure 3.** (a) Arrhenius plot of the experimental relaxation time of the molecular dipolar relaxations (I and II) in the solid (α) phase of TCE, compared to simulation results. Filled circles and triangles represents experimental relaxation times obtained from the fit of the imaginary permittivity, while open triangles those corresponding to the fit of the modulus spectra. Empty

squares correspond to the two decay times extracted by the double exponential fit of the self-correlation functions at $T_m$ – 20 K, $T_m$ – 10 K and ~ $T_m$, respectively. Large filled circles with error bars are typical waiting times for molecular jumps between equilibrium configurations (see the text for more details), at the same simulation temperatures. Inset: temperature dependence of the dielectric strength of relaxation I, as determined from the permittivity fit. (b) Self-correlation functions of the molecular dipole vector (Eq. 4) at the simulation temperature of $T_m$ – 20 K (green dotted curve), and of the bond orientations (Eq. 5) for two of the C–Cl bonds (involving different carbon atoms) of the same molecule (blue dashed and orange dot dashed curves) at $T_m$ – 20 K, $T_m$ – 10 K, and ~$T_m$. Continuous lines are fits with a double exponential function.

The temperature dependence of the relaxation times of process I could be modeled with a simply activated (Arrhenius) behavior, with activation energy of 39.3 kJ/mol and high-temperature relaxation time $\tau_\infty = 0.68 \cdot 10^{-14}$ s, which is consistent with a dipolar dynamics.[43] Process II displayed a more pronounced dependence on temperature, well described by the Vogel-Fulcher-Tamman (VFT) equation, which is given by:

(Eq. 6) $\tau_{max} = \tau_\infty \exp[D\, T_{VF}/(T - T_{VF})]$.

Here the prefactor $\tau_\infty$, the fragility parameter $D$ and the so-called Vogel-Fulcher temperature $T_{VF}$ are phenomenological parameters, with values $\tau_\infty = 6.31 \cdot 10^{-12}$ s, $D = 11$, and $T_{VF} = 87.2$ K). The fit of the relaxation times with an Arrhenius equation or with Eq. (6) also allows determining the vitrification point of both processes, defined as the temperature at which their relaxation time approaches 100 s (see dotted horizontal line in Figure 3(a)) – in other words, the vitrification temperature is defined as to coincide with the glass transition temperature ($T_g$) if the processes were the main cooperative dynamics in a glass-forming system. The estimated vitrification temperatures are 127 ± 1 K and 119 ± 1 K for process I and II, respectively. It is interesting to note that the two vitrification temperatures are not far from one another.

The behavior of relaxation II is the expected one for a cooperative reorientational process: it is observed at all temperatures, displays VFT behavior and a dielectric strength that decreases slowly with temperature. In this sense, this relaxation is the intrinsic relaxation of the α phase, and it is reminiscent of the observed dynamics in the solid phase of another halogenated ethane derivative, (1,1,2,2)-tetrachloroethane.[25,44] On the other hand, process I has, as mentioned, a strength that increases strongly with increasing temperature. Moreover, the disorder with which it is associated is absent far below the melting point. In this respect, process I has features that make it unique to the solid phase of TCE.

To compare experiment with simulation, we carried out a statistical analysis, in terms of self-correlation functions, of our molecular dynamic simulation data. The time self-correlation function of the single-molecule dipole moment $p$ (Eq. (4)) is shown with a dotted line in Figure 3(b). Since the decay of the correlation is too slow in time to allow a reliable analysis, we focus on the C–Cl bond correlation defined by Eq. (5), which is shown for one C–Cl bond of each carbon atom of a TCE molecule in the same Figure 3(b), at three different temperatures. For each temperature, the bond self-correlation function was fitted as the sum of two exponential functions, with the same time constant for all C–Cl bonds of the same molecule. The existence of two decay constants is indicative of the presence of two main dynamic ranges, one at lower frequency, corresponding to molecular dynamic processes that occur more seldom, and the other at higher frequency, corresponding to processes that occur more often. The obtained characteristic self-correlation times are shown as empty squares in Figure 3(a), to allow direct comparison with the dielectric data. It can be seen that, for relaxation I, a good agreement is found between the experimental and simulated correlation times, also with similar temperature dependence, while the extrapolation of relaxation II to high temperature gives values that are compatible with the simulation results. Hence, both experiment and simulation show the existence of two main characteristic times of molecular dynamics.

Since dielectric spectroscopy measures the total polarization of a sample, processes I and II actually represent the average, macroscopic relaxations of the TCE molecules as a whole. In order to gain insight on the microscopic origin of the two relaxation processes, we analyzed in detail the motion of individual molecules during our molecular dynamics simulation runs. The local dynamics taking place in the simulations involve jump-like transition between distinct configurations of a given molecule. One can observe several types of molecular jumps, which can be classified into two main groups depending on the average time-duration of the molecular configuration reached through such jumps. The first group of motions involves leaps between different equilibrium configurations of a molecule; after such a leap, the individual molecule stays in the final state for a relatively long time, generally of the order of hundreds or even thousands of picoseconds (ps). The second group of motions consists of fluctuations in which the molecule only stays for at most tens of ps in an intermediate configuration, to return then to an equilibrium one.

As an example, the top panel of Figure 4 shows the coordinates of each chlorine atom of a representative molecule, as a function of simulation time. In the top left part, it can be observed that the chlorine positions and thus the average molecular conformation remains fixed during thousands of ps. As visible in the two molecular cartoons shown in the bottom left corner, these relatively stable "equilibrium" configurations of a given molecule always display the chlorine atoms in the same initial positions. The direct transitions between two equilibrium configurations involve a large-angle variation of the direction of the C–C bond, and a simultaneous conformational leap between *gauche*⁺ and *gauche*⁻ isomers. The conformational change is required precisely to maintain the invariant the position of the chlorines. These transitions correspond to the dynamics suggested by the authors of Ref. 14 to account for the observed structural disorder close to the melting point. Our simulations reveal that the simultaneous orientation-configuration jumps actually take place through two mechanisms: in one of them (bottom left cartoon) all three chlorine atoms are interchanged, as visible from the labeling of the individual chlorines; in the other (cartoon next to the previous one), one of the chlorine atoms remains in its initial position while the other two are interchanged.

By enlarging the horizontal (time) scale, it can be observed that the transition between the long-lived states is not always direct, but can also involve short-time fluctuations (few ps) into molecular configurations in which some of the chlorine atoms occupy positions that are different from the equilibrium ones. Such fluctuations occur also during the longer periods where the molecule appears to be always in the same configuration; in fact, it turns out that the fluctuations to a non-equilibrium orientation are more frequent than the jumps between equilibrium states (see Figure 5(a)). While the latter always involve a simultaneous orientational and conformational leap, the short-time fluctuations may or may not involve a change of conformation. Just as there are two distinct types of jumps between equilibrium configurations, there exists several possible fluctuations; two of them are depicted in the bottom right corner of Figure 4. Some fluctuations involve the interchange of two chlorine atoms while the third one occupies a non-equilibrium position; in some of them, the out-of-equilibrium configuration has two chlorine atoms that are out of place. Some of the fluctuations can be best described as large-amplitude librations, with no conformational changes.

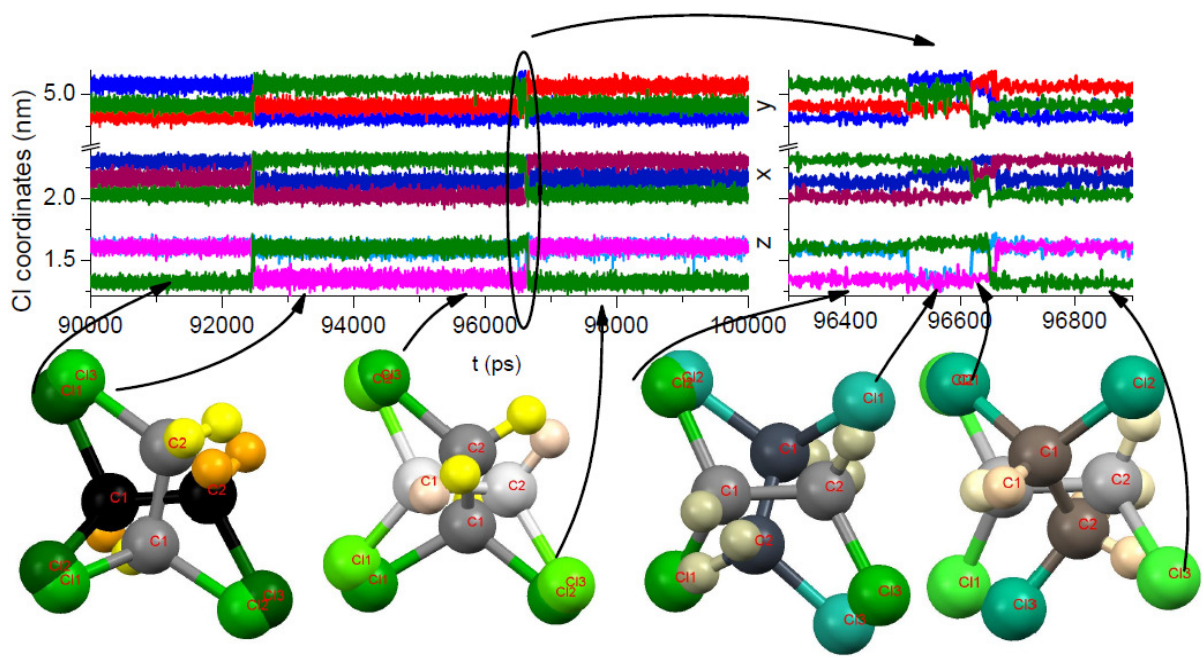

**Figure 4**. Top panel: (x,y,z) coordinates of the three chlorine atoms of a representative TCE molecule during a simulation run at $T \approx T_m$, as a function of simulation time. The right-hand side shows a detail made visible by enlarging the horizontal (time) scale. Bottom panel: cartoons depicting the configurational changes undergone by the molecule at different times during the simulation run. The two cartoons on the left represent jumps between equilibrium configurations, the two on the right, fluctuations involving short-lived (ps) molecular configurations with one chlorine atom completely out of its equilibrium position.

It is natural to assign the observation of two clearly separate relaxation times to the presence of two distinct types of dynamics. In view of the similarity of experimental and simulation relaxation times, and in view of the fact that the jumps between equilibrium configurations involve a simultaneous orientational and conformational change, just as required for the dynamic disorder of Figure 1 and observed by x-ray diffraction,[14] we assign dielectric relaxation I to the interconversion dynamics between equilibrium conformers, such as that depicted in the left-hand side of Figure 4. This assignment is corroborated also by the analysis of the mean waiting time between successive equilibrium jumps: as we did in a previous work,[45] we have estimated the mean time between jumps by counting the number of jumps within a 100 ns simulation interval at a given simulation temperature. The results for the temperatures of ~$T_m$, $T_m - 10$ K, $T_m - 20$ K and $T_m - 30$ K are shown as filled circles with error bars in Figure 3(a). It can be observed that in all cases the waiting time matches perfectly the longer self-correlation time, thus showing that the longer time constant (corresponding to the experimental relaxation I) is due to simultaneous conformational/orientational interconversion jumps. All these results match perfectly those obtained from the detailed structural analysis reported Ref. 14, and they moreover provide hitherto unknown details concerning the intrinsic molecular dynamics associated with the peculiar site disorder in solid TCE.

Concerning the fluctuations, their short duration and the fact that they occur more frequently than the "equilibrium jumps" indicate that their corresponding relaxation times should be much shorter than those of relaxation I. We therefore assign relaxation II to short-lived fluctuations, such as those depicted in the bottom right part of Figure 4. It should be pointed out that these fluctuations represent hitherto unknown molecular dynamic process taking place in solid TCE. The short life-time of the intermediate state reached during a fluctuation rationalizes the fact that no significant distortion is detected by XRD analysis.[14] In this sense, dielectric spectroscopy and molecular dynamics simulations are much more powerful than XRD (which measures only time-average structures) in detecting molecular dynamics of apparently ordered solids.[46-52]

Since the relaxation processes are related to variations of the dipole moments of the individual molecules, it is worth analyzing the local molecular dynamics in terms of the single-molecule dipole moments. When a jump takes place between equilibrium molecular configurations, the direction of the C–C bond changes by about 60$^{\text{o}}$ and those of the C–Cl bonds between 90$^{\text{o}}$ and 150$^{\text{o}}$. By contrast, the molecular dipole vector changes only by 30$^{\text{o}}$ its orientation. On the other hand, several fluctuations lead to short-lived non-equilibrium states whereby the C–C direction changes by a similar angle of 60$^{\text{o}}$, but under which the molecular dipole moment changes by a very large angle, even close to 180$^{\text{o}}$. This can be observed in Figure 5, which depicts the dynamics of a molecule undergoing several fluctuations. In panels (a) and (b), the positions of the chlorines are shown as a function of simulation time, while panels (c) and (d) display the end position of the molecular dipole vector at different instants during the simulation run.

Four time intervals can be identified in panel (a), labeled from *i* to *iv*. It can be seen that, in each time interval, the chlorines occupy most of the time equilibrium positions, corresponding to a stable molecular configuration, and that in between these comparatively long intervals, a configurational leap takes place (corresponding to relaxation I). From the chlorine coordinates it may be inferred that the jumps separating intervals *i* and *ii* (46300 ps) and *iii* and *iv* (62800 ps) both occur by interchange of two out of three chlorines, while the third one remains unchanged; instead, the jump between *ii* and *iii* (49700 ps) takes place by interchange of all three chlorines. Within a given interval, especially in interval *iii*, several "spikes" can be observed in the chlorine coordinates, each corresponding to a short-lived fluctuation. Several "fluctuations" in the interval *iii* (see panel (b) of Figure 5) actually correspond to relatively long-lived configurations (100 ps) where two out of three chlorines are in an out-of-equilibrium position. On the other hand, in time intervals *ii* and *iv* mainly short-lived (few ps) fluctuations take place. The end position of the molecular dipole during each time interval is shown with a different color in Figures 5(c,d). It is clear that, while in all time intervals the dipole vector points most of the time in roughly the same direction, which differs only by a relatively small angle (30$^{\text{o}}$) in each equilibrium configuration, large-angle fluctuations of the dipole orientation take place, especially in the time interval *iii*.

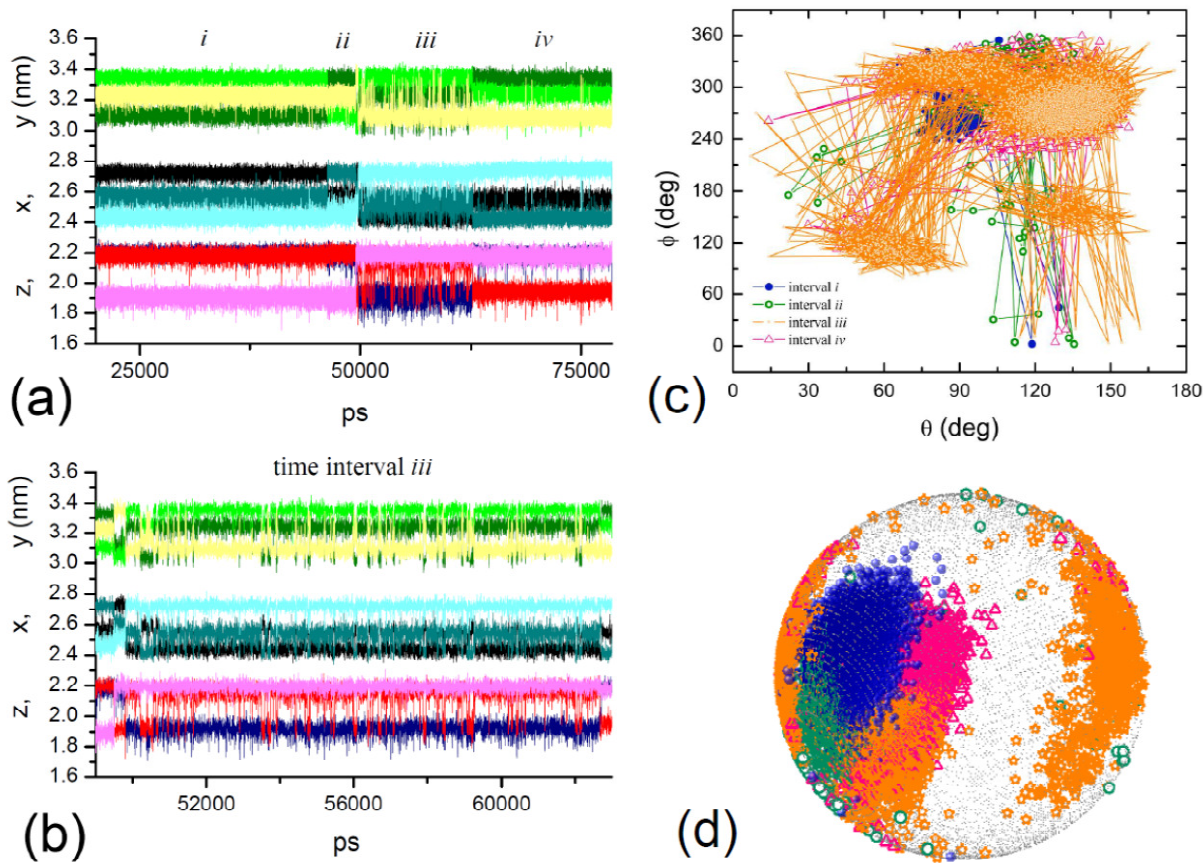

**Figure 5.** (a,b) Coordinates of the three chlorine atoms of a TCE molecule undergoing several fluctuations, as a function of simulation time at $T_m - 10$ K. Four separate time intervals are identified, labeled from *i* to *iv*. Panel (b) is a zoom-in of time interval *iii*. (c,d) Orientation of the molecular dipole during all four time intervals, depicted in two representations: as planar plot as a function of the spherical coordinate angles (c), and as spherical projection of the trajectory of the dipole vector (d).

It should be remarked that, according to our molecular dynamics and the consequent interpretation of dielectric spectroscopy results, both macroscopic relaxations (I and II) actually correspond to more than one possible microscopic (molecular) dynamics: as exemplified in Figure 4 and 5, a plethora of local processes are observed in which the direction of the dipole moment changes either by small or large angles; in which the initial and final positions of the all three chlorine atoms are maintained, or only those of two chlorines or a single chlorine atom; in which the molecular conformation stays the same, as in a libration, or else changes to a different chirality, as in a conformational conversion; and in which the final state is an equilibrium state or else a short-lived non-equilibrium configuration where the final chlorine positions do not coincide with crystallographic allowed sites. The fact that the dipole vector changes only few degrees with each configuration jump (relaxation I), and the fact that it stays only for a very short time in the non-equilibrium state during a fluctuation (relaxation II), both rationalize the weak time dependence of the dipole self-correlation function (Figure 3(b)).

Concerning the dielectric strength of each relaxation process, in principle it reflects the ensemble average polarization changes of the sample due to a given molecular dynamics, and it depends on several factors such as the number of molecules involved, the waiting time, and the mutual correlation between molecular leaps. In a solid phase such as the α phase of TCE, all dynamics are likely to have an at least partial cooperative character due to the strong intermolecular cohesive interactions. The VFT temperature dependence of process II indeed indicates that the fluctuations display cooperative character, possibly associated with the fact that steric interactions are enhanced when the chlorine species occupy non-equilibrium positions. The size of cooperatively rearranging regions, or in other words the influence of a molecular leap on the surrounding molecules, is a subject of debate even in "canonical" glasses.[53] "Well-ordered" solids like the one studied here, in which correlation functions are fully defined and in which the compatibility between structural requirements and dynamics provide additional physical constraints, represent interesting model systems to further investigate these open questions.

Conclusions

We employ molecular dynamics simulations and dielectric spectroscopy to study the ambient-pressure stable crystal phase of (1,1,2)-trichloroethane (TCE). Our simulations show that both *gauche*$^+$ and *gauche*$^-$ conformers are present in this solid phase, occupying preferently specific crystallographic sites. While the total population of each conformer phase α is always 50%, independent of temperature, between the melting point $T_m$ and $T_m - 40$ K a considerable fraction of crystallographic sites are occupied by the minority conformer, in agreement with previous x-ray diffraction results (Ref. 14). The onset of this disorder is signaled by an anomalous temperature dependence of the (ionic) dc conductivity and of the space-charge relaxation frequency in a ten degree range around approximately $T_m - 30$ K.

Two macroscopic relaxations of the sample's polarization are observed by means of dielectric spectroscopy, which stem from distinct molecular dynamic processes. Our MD simulations show that the slower molecular dynamics (relaxation I) involves a simultaneous reorientational and conformational change of the TCE molecules, namely between *gauche*$^+$ and *gauche*$^-$ conformers of well-defined and distinct orientation. The observation of this process both in dielectric experiments and molecular simulations confirms the idea of Ref. 14 that the disorder in solid TCE is dynamic in nature and involves simultaneous orientational/conformational leaps. Our simulations also unveil the existence of short-lived (few ps) molecular fluctuations that are undetectable by x-ray diffraction, and whose existence and frequency are able to account for the experimental observation of the faster molecular relaxation process (relaxation II). The simulation results indicate that such macroscopic relaxation stems both from conformational fluctuations and large-angle librations of the TCE molecules, which are short-lived because the intermediate state in both cases is a non-equilibrium, high-energy configuration with enhanced steric repulsion between the chlorine atoms of next-neighbor molecules.

Our study confirms once more the vast richness of molecular dynamics of ethane derivatives in the solid state. In particular, simultaneous orientational-conformational dynamics are quite rare in solid phases, and it is therefore interesting that the TCE molecules exhibit more than one local dynamics where both changes take place simultaneously. Our results show that the combination of molecular dynamics simulation with a sensitive probe such as dielectric spectroscopy is required to fully unravel the dynamic disorder of small-molecule organic solids.

AUTHOR INFORMATION

**Corresponding Author**

* Author to whom correspondence should be addressed. Electronic mail:

roberto.macovez@upc.edu

**Present Addresses**

† Current address: Laboratory of Polymers and Soft Matter Dynamics, Faculté des Sciences, Université libre de Bruxelles (ULB), CP 223, Avenue Franklin Roosevelt 50, 1050, Ixelles, Brussels, Belgium.

**Author Contributions**

The manuscript was written through contributions of all authors.

**Notes**

The authors declare no competing financial interests.

ACKNOWLEDGMENT

This work has been partially supported by the Spanish Ministry of Economy and Competitiveness MINECO through project FIS2014-54734-P and by the Generalitat de Catalunya under project 2014 SGR-581. M.Z. and P.S. acknowledge financial support of the Argentinian SECYTUNC and CONICET. This work used computational resources from CCAD Universidad Nacional de Córdoba (http://ccad.unc.edu.ar/), in particular the Mendieta Cluster, which is part of SNCAD MinCyT, República Argentina.

ABBREVIATIONS

TCE; (1,1,2)-trichloroethane; HN, Havriliak Negami; CC, Cole-Cole; CD, Cole-Davidson; VFT, Vogel-Fulcher-Tamman.